\documentclass{aa}  

\usepackage{graphicx}
\usepackage{txfonts}
\def\ltsima{$\; \buildrel < \over \sim \;$}
\def\simlt{\lower.5ex\hbox{\ltsima}}
\def\gtsima{$\; \buildrel > \over \sim \;$}
\def\simgt{\lower.5ex\hbox{\gtsima}}
\def\gsimeq
{\hbox{\raise0.5ex\hbox{$>\lower1.06ex\hbox{$\kern-1.07em{\sim}$}$}}}
\def\lsimeq
{\hbox{\raise0.5ex\hbox{$<\lower1.06ex\hbox{$\kern-1.07em{\sim}$}$}}}
\def\pn{\par\noindent}
\def\ss{\smallskip\pn}

\def\xmm{{\it XMM-Newton }}
\def\rosat{{\it ROSAT}}

\def\xmm{{\it XMM-Newton}}

\def\fermi{{\it Fermi}}

\def\meerkat{{\it MeerKAT}}
\def\herschel{{\it Herschel}}
\def\wise{{\it WISE}}
\def\msx{{\it MSX}}

\def\iras{{\it IRAS}}
\def\athena{{\it Athena}}

\def\apj{ApJ}
\def\aj{AJ}
\def\mnras{MNRAS}
\def\aap{A\&A}

\def\apjl{ApJ}
\def\apjs{ApJS}
\def\araa{ARA\&A}
\def\pasj{PASJ}
\def\nat{Nature}

\def\af{AFGL~5376}

\def\sgras{Sgr~A$^\star$}

\begin{document} 

   \title{The Galactic center chimneys:\\
   The base of the multiphase outflow of the Milky Way}

   \author{G. Ponti\inst{1,2} 
   \and
   M. R. Morris\inst{3}
   \and
   E. Churazov\inst{4,5}
   \and
   I. Heywood\inst{6,7,8}
   \and
   R. P. Fender\inst{6}
   }
   \institute{INAF-Osservatorio Astronomico di Brera, Via E. Bianchi 46, I-23807 Merate (LC), Italy \\
         \email{gabriele.ponti@inaf.it}
         \and
         Max-Planck-Institut f{\"u}r Extraterrestrische Physik, Giessenbachstrasse, D-85748, Garching, Germany  
         \and
        Department of Physics and Astronomy, University of California, Los Angeles, CA 90095-1547, USA 
        \and
        Max-Planck-Institut fur Astrophysik, Karl-Schwarzschild-Str. 1, D-85748, Garching, Germany
        \and
        Space Research Institute (IKI), Profsoyuznaya 84/32, Moscow 117997, Russia
        \and
        Astrophysics, Department of Physics, University of Oxford, Keble Road, Oxford OX1 3RH, UK
        \and
        Department of Physics and Electronics, Rhodes University, PO Box 94, Makhanda, 6140, South Africa
        \and
        South African Radio Astronomy Observatory, 2 Fir Street, Black River Park, Observatory, Cape Town, 7925, South Africa
             }
   \date{Received October 9, 2020; accepted December 25, 2020}

  \abstract
   {Outflows and feedback are key ingredients of galaxy evolution. Evidence for an outflow arising from the Galactic center (GC) -- the so-called GC chimneys -- has recently been discovered at radio, infrared, and X-ray bands. }
   {We undertake a detailed examination of the spatial relationships between the emission in the different bands in order to place constraints on the nature and history of the chimneys and to better understand their impact on the GC environment and their relation with Galactic scale outflows. } 
   {We compare X-ray, radio, and infrared maps of the central few square degrees.} 
   {The X-ray, radio, and infrared emissions are deeply interconnected, affecting one another and forming coherent features on scales of hundreds of parsecs, therefore indicating a common physical link associated with the GC outflow. We debate the location of the northern chimney and suggest that it might be located on the front side of the GC because of a significant tilt of the chimneys toward us. 
   We report the presence of strong shocks at the interface between the chimneys and the interstellar medium (ISM), which are traced by radio and warm dust emission. We observe entrained molecular gas outflowing within the chimneys, revealing the multiphase nature of the outflow. In particular, the molecular outflow produces a long, strong, and structured shock along the northwestern wall of the chimney. Because of the different dynamical times of the various components of the outflow, the chimneys appear to be shaped by directed large-scale winds launched at different epochs. 
   
   The data support the idea that the chimneys are embedded in an (often dominant) vertical magnetic field, which likely diverges with increasing latitude. We observe that the thermal pressure associated with the hot plasma appears to be smaller than the ram pressure of the molecular outflow and the magnetic pressure. This leaves open the possibility that either the main driver of the outflow is more powerful than the observed hot plasma, or the chimneys represent a "relic" of past and more powerful activity. }
   {These multiwavelength observations corroborate the idea that the chimneys represent the channel connecting the quasi-continuous, but intermittent, activity at the GC with the base of the \fermi\ bubbles. In particular, the prominent edges and shocks observed in the radio and mid-infrared bands testify to the most powerful, more recent outflows from the central parsecs of the Milky Way. }


   \maketitle
%

\section{Introduction}

Outflows and feedback are vital ingredients for the forming and growing of galaxies as we observe them today. Outflows are required in order to connect the activity in the cores and disks of galaxies with the hot, slowly recondensing plasma in their haloes, thereby fostering the evolution of galaxy morphologies (White et al. 1978; 1991; Putman et al. 2012; Tumlinson et al. 2017). Such feedback links the growth of the central supermassive black holes with their coevolving galaxy (Ferrarese \& Merrit 2000; Gebhardt et al. 2000; Kauffmann et al. 2003). 

As a prototype for typical spiral galaxies, the Milky Way offers a unique opportunity to capture the important details of such feedback all the way from sub-parsec to galactic scales. Indeed, as the Milky Way is located at a distance of only $8.25$~kpc (Gravity Col 2020; see also Do et al. 2019, who suggested $\sim7.97$~kpc), we can investigate its physical processes at a resolution orders of magnitudes better than in other quiescent galaxies. The most pressing outstanding question is how some portion of the multiphase interstellar medium (ISM) can be launched from galactic centers and disks into outflows that replenish galactic coronae, haloes, or even the intergalactic medium with plasma, energy, metals, etc. (Naab et al. 2017). This involves understanding the complex physics of galaxies and their multiphase gas. 
\begin{figure*}
\begin{center}
\includegraphics[width=1.3\textwidth,angle=-90]{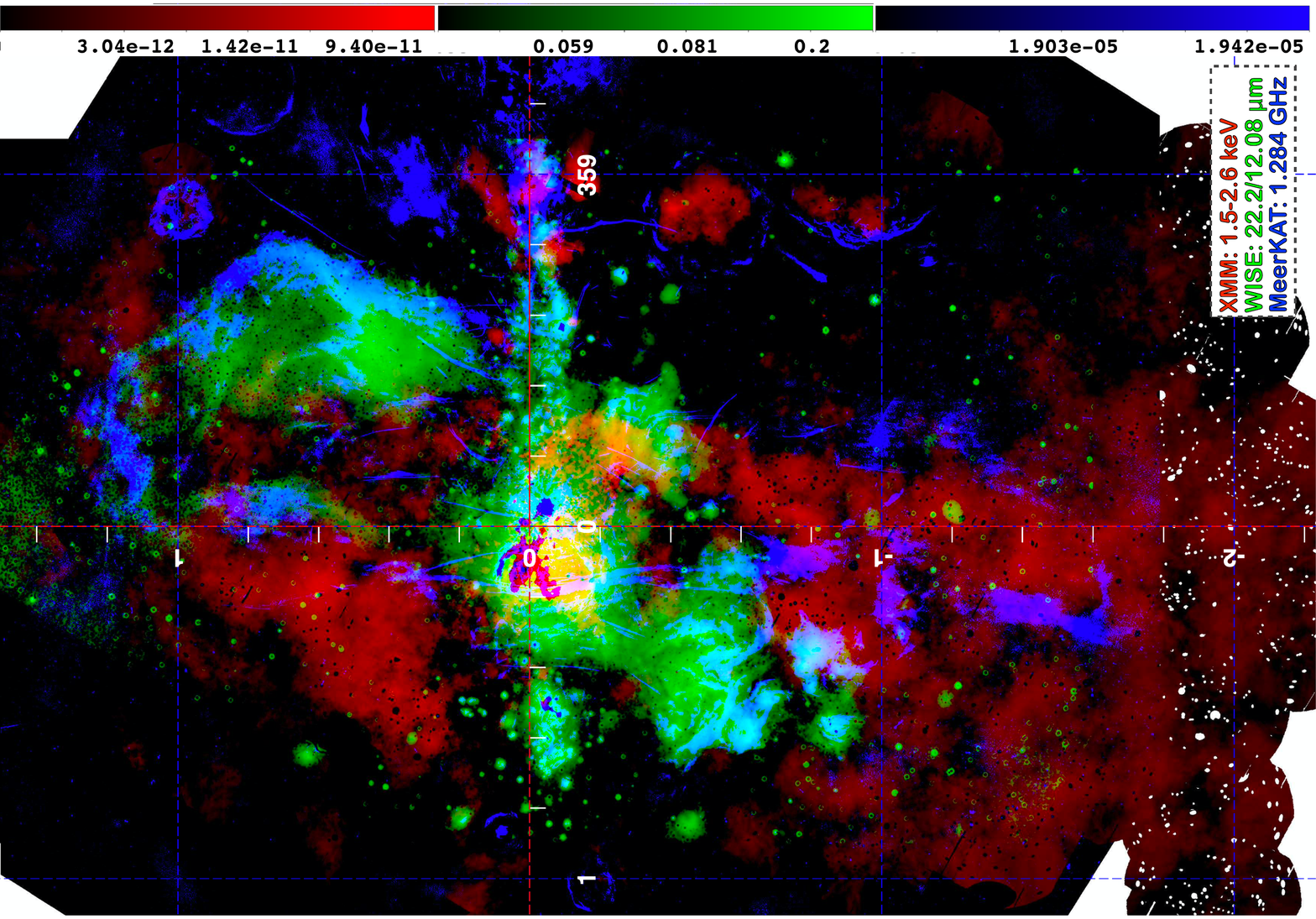}
\vspace{-0.4cm}
\end{center}
\caption{Multi-phase nature of the GC chimneys. Red, green, and blue show the \xmm\ 1.5-2.6~keV, \wise\ (ratio of 22.2~$\mu$m/12.08~$\mu$m emission), and \meerkat\ 1.284 GHz maps, respectively. }
\label{all}
\end{figure*}

The detection of hints of an outflow from the Galactic center (GC) dates back to the 1980s, when sensitive radio maps revealed features with an extent of a few degrees (such as the so-called expanding molecular ring and the Galactic center lobe; GCL), which were originally attributed to large energy releases from the core of the Milky Way (Kaifu et al. 1972; Scoville 1972; Sofue 1984; 1985; 1989). Subsequently, the combination of X-ray (\rosat) and mid-infrared (\iras\ and \msx) observations strengthened this hypothesis, revealing a limb-brightened bipolar structure, possibly the outcome of a large-scale bipolar wind from the GC (Bland-Hawthorn \& Cohen 2003). 
This scenario was then brought to the fore by the discovery of the so-called \fermi\ bubbles, clearly visible above $\sim2$~GeV in the \fermi\ data, with a size comparable to the Milky Way itself and a total energy content of $\sim10^{55}$~erg (Su et al. 2010; Ackermann et al. 2014; Kataoka et al. 2018). It was also suggested that the bases of the \fermi\ bubbles are associated with soft X-ray emission (Bland-Hawthorn \& Cohen 2003; Su et al. 2010; Nakashima et al. 2013; Crocker et al. 2015).

Recently, we reported sensitive X-ray maps of the GC, which led us to the discovery of two oppositely directed, 200-pc chimneys of hot plasma connecting the central parsecs with the base of the \fermi\ bubbles (Ponti et al. 2015; 2019; Nakashima et al. 2019). Such chimneys are the smoking-gun evidence of an outflow from the GC (Ponti et al. 2019). Subsequently, radio maps revealed extended radio continuum emission defining two edge-brightened lobes or bubbles, roughly tracing the edges of the X-ray chimneys (Heywood et al. 2019). 

Here we examine the X-ray maps jointly with the radio and infrared maps. 
In Sect. 2, we describe the overlay of the X-ray maps with the radio and infrared ones. In Sect. 3, we discuss the results and then consolidate the observed complexity. Finally, Sect. 4 proposes an emerging simplified picture and details our conclusions. 
\begin{figure*}
\begin{center}
\vspace{-0.4cm}
\includegraphics[width=1.14\textwidth,angle=-90]{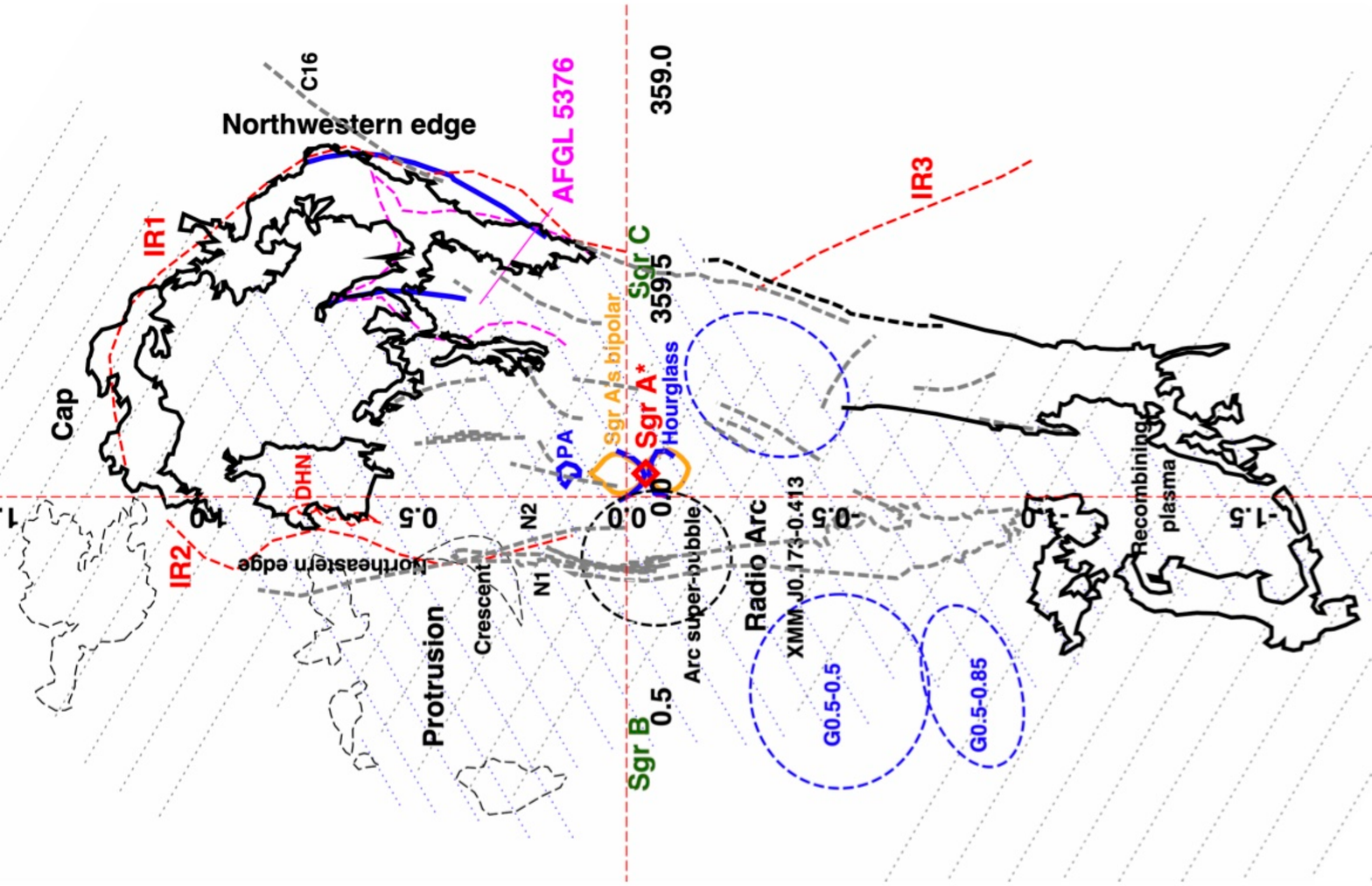}
\vspace{-0.5cm}
\end{center}
\caption{Finding chart of the region within and just outside the chimneys. The thick-black solid regions show the most prominent radio emitting features clearly associated with the chimneys. The thick dashed gray lines display the locations of the most prominent nonthermal radio filaments (such as N1, N2, XMMJ 0.173-0.413, C16) and of the Radio Arc. The thick-red dashed lines display the location of the spurs of warm dust as evinced from the \wise\ maps and the double helix nebula (DHN; Bland-Hawthorn \& Cohen 2003; Enokiya et al. 2014). The magenta dashed lines indicate the borders of the \af\ feature, as derived from the \wise\ 22.2~$\mu$m map (the solid blue lines at this location show the position of the two shocks derived by Uchida et al. 1994). 
The slanting thin-dotted blue and gray lines show the X-ray emitting region displaying bright S~{\rm xv} and soft X-ray line radiation, respectively. At high latitudes, this emission extends beyond the borders of the chimneys. 
The thin-black dashed regions show the location of faint radio features appearing at the edges of the X-ray protrusion (\S \ref{protrusion}). At the center, a red diamond shows the location of \sgras, the orange solid region displays the location of Sgr~A's bipolar lobes, as derived from the \xmm\ map, while the blue solid regions show the location of the polar arc (PA) and of the hourglass feature (Hsieh et al. 2016). The latter runs almost perfectly on top of the lower edges of Sgr~A's bipolar lobes. The thick-black dashed ellipse shows the location of the Arc super-bubble as it appears in the radio and mid infrared bands, while the thick-blue dashed ellipses show the location of the three well-known foreground star forming regions. The dark green labels indicate the position of the main molecular complexes (the Sgr A complex approximately coincides with \sgras).}  
\label{FC}
\end{figure*}

\section{Multiwavelength view of the multiphase GC outflow}

The interplay between the various phases of the GC outflow is shown in Figure \ref{all}. The red, green, and blue colors show the X-ray (\xmm), infrared color ratio (22.2~$\mu$m/12.08~$\mu$m from \wise), and radio (\meerkat) maps, respectively. In red, the continuum-subtracted 1.5-2.6~keV map is shown in logarithmic scale (see Extended Data Fig. 3 of Ponti et al. 2019). Because the X-ray chimneys are primarily thermally emitting, they shine brightly in soft X-ray emission lines. The green shows the ratio of infrared color defined as the 22.2~$\mu$m \wise\ map divided by the 12.8~$\mu$m one (see \S~\ref{secwise} for more details). In blue, the \meerkat\ map is shown at intensities at or above $2\times10^{-5}$ Jy (see Fig. 1 of Heywood et al. 2019 and note the caveats regarding photometric accuracy in the Methods section). This outstanding color image shows the interplay of the different phases of the GC outflow. 
 
We also note that the \meerkat\ and \wise\ maps show bright radio and infrared emission associated with G0.5-0.5 and G0.5-0.85, two well-known foreground star formation complexes. 
\begin{figure*}
\begin{center}
\includegraphics[width=1.3\textwidth,angle=-90]{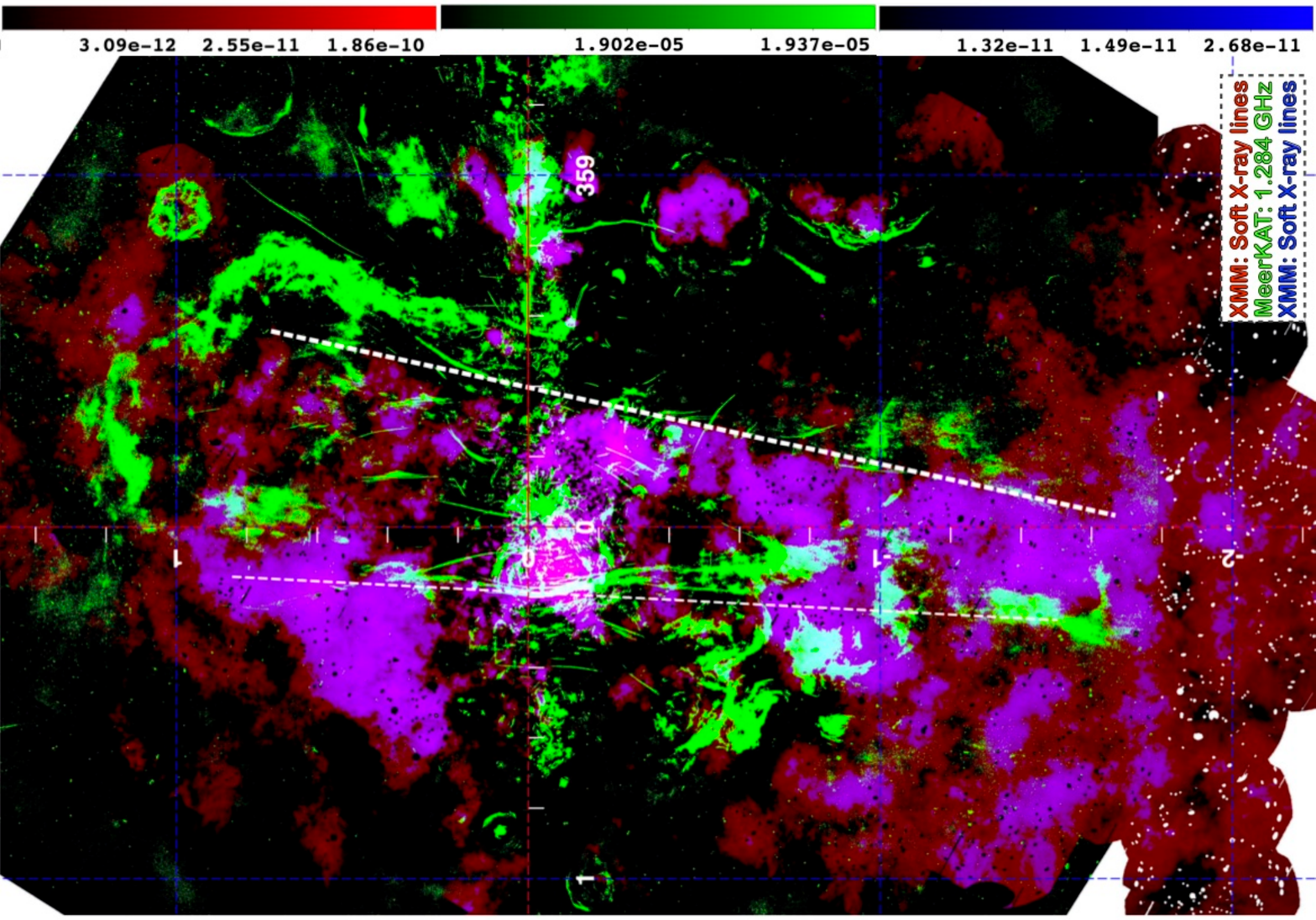}
\end{center}
\caption{Continuum-subtracted soft X-ray line image in red and blue (with different intensity cuts, to give a better indication of the extent of the chimneys). In green, the \meerkat\ map with intensity cut at $2\times10^{-5}$ Jy (see Fig. 1 of Heywood et al. 2019, and note the caveats regarding photometric accuracy in the Methods section). }
\label{xmmMK}
\end{figure*}
Figure \ref{FC} shows a finding chart of the region within and just outside of the chimneys, displaying most of the features discussed here. 

\subsection{Overlay of radio and X-ray maps}

\begin{figure*}
\begin{center}
\includegraphics[width=1.3\textwidth,angle=-90]{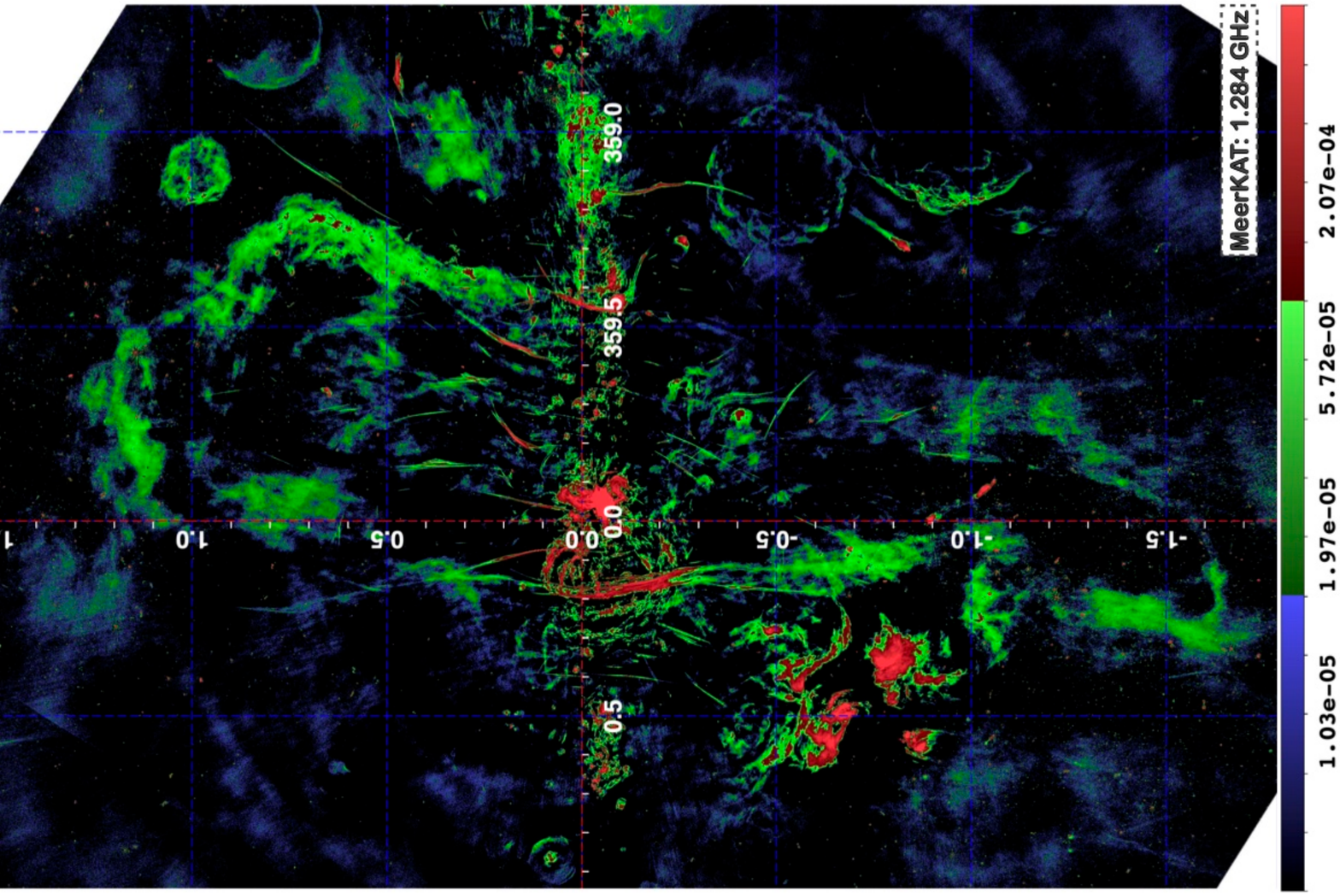}
\end{center}
\caption{\meerkat\ map displayed with its entire dynamical range.}
\label{MeerKAT}
\end{figure*}
Figure \ref{xmmMK} shows an X-ray (\xmm) and radio (\meerkat) overlay. The thick dashed white line indicates the location of the prominent western edge of the distribution of X-ray emitting plasma defining the chimneys. This appears as a remarkably linear feature with an extent of $\sim350$~pc. The thin dashed white line aims at indicating a possible location of the eastern edge of the chimneys; however, its position is less well determined.

We note that both the northern and southern chimneys are prominent in both the X-ray and radio bands, with a striking degree of symmetry with respect to the Galactic plane. However, some asymmetries are clearly evident (Fig. \ref{xmmMK}). 

Toward the northern latitudes, the X-ray emission associated with the GC outflow appears consistent with being edge-brightened (i.e., lacking a ridge of emission at the center of the chimney), although not as much as the radio emission (e.g., see Additional Data Figure 7 of Ponti et al. 2019). This might indicate that the X-ray emission is also produced primarily at the boundaries of the GC outflow, possibly in a structured shock with the ISM. On the contrary, the X-ray emission toward the southern hemisphere peaks along the axis of the chimney, as would be expected if the X-ray emitting plasma is volume filling. 

Figure \ref{MeerKAT} shows the full dynamic range of the \meerkat\ map, which displays a large array of nonthermal filaments and diffuse radio emission. Some of these filaments are reported in Fig. \ref{FC} as dashed gray lines. 

\subsection{Maps of X-rays versus dense, neutral material}
\begin{figure*}
\hspace{-0.4cm}
\includegraphics[width=1.05\textwidth,angle=0]{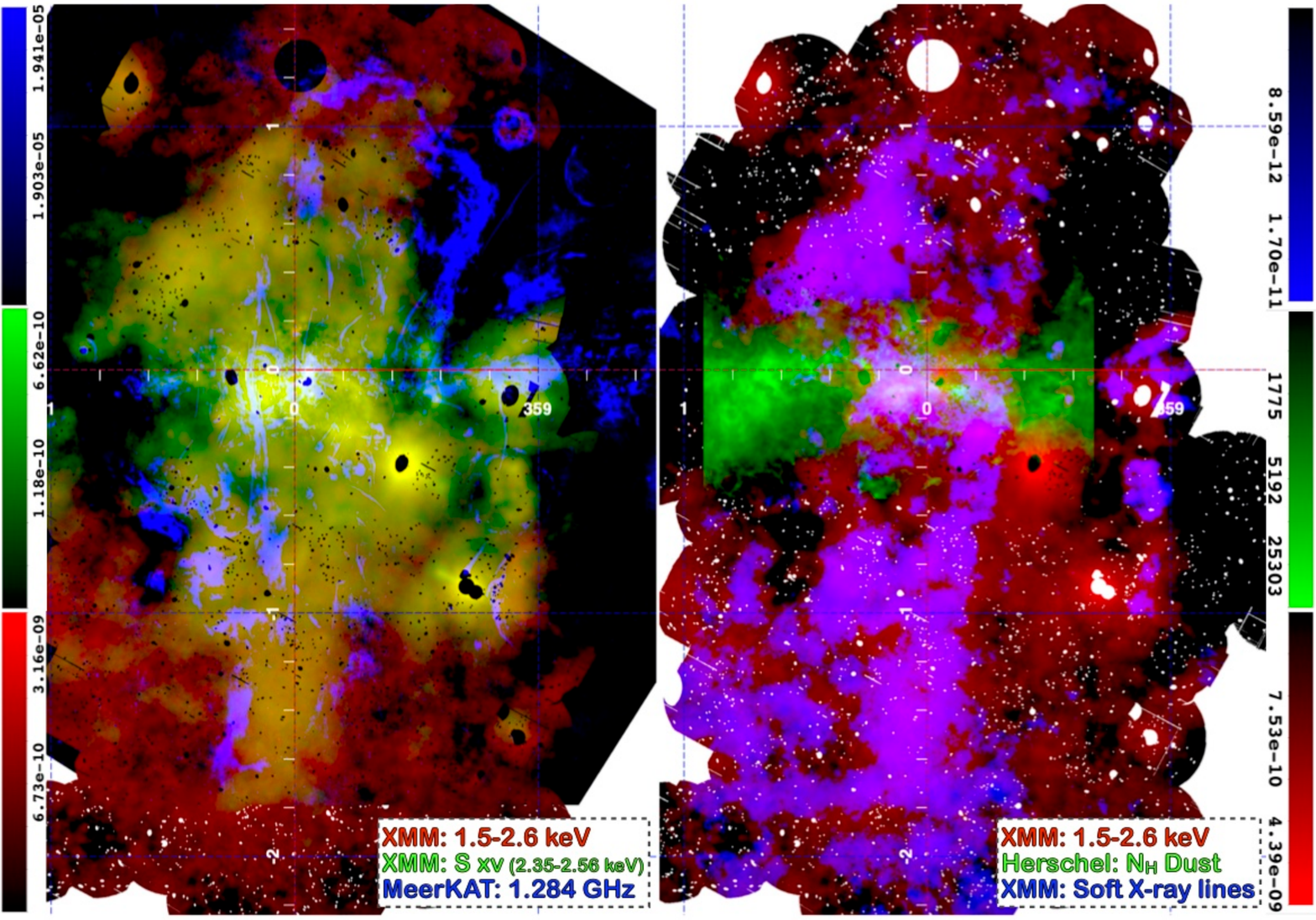}
\caption{Multiwavelength view of the chimneys. {\it (Left panel)} Red colors show the soft X-ray emission map (1.5-2.6 keV). Green colors show the S~{\sc xv} line emission (2.35-2.56~keV) and blue colors show the \meerkat\ map. The S~{\sc xv} line emission shows clear gradients in rough coincidence with the edges of the \meerkat\ bubbles. Additionally, it is less affected by interstellar absorption than the soft X-ray emission map (indeed the latter shows clear depressions in the emissivity of the chimneys at $b\sim0.2^\circ$ and $b\sim-0.4-0.6^\circ$). 
{\it (Right panel)} RGB image of the GC. Red: Soft X-ray emission map (1.5-2.6 keV). Green: Column density map of neutral material as derived from dust emission of the central molecular zone, derived from \herschel\ observations (Molinari et al. 2011). The depressions of soft X-ray emission at $b\sim0.2^\circ$ and $b\sim-0.4-0.6^\circ$ correspond to a high column density of neutral material, probably as a result of enhanced obscuration. We point out that the protrusion at $l=0.2$, $b=0.25^\circ$ (\S \ref{protrusion}) might be associated with a fountain originating from the Sgr B molecular complex, one of the most massive, highly obscured and highest star formation rate locations of the Milky Way. Indeed, although the protrusion appears to be disconnected from the plane in the soft X-ray map (it extends from $b\sim0.2^\circ$ to $b\sim0.6^\circ$), this might be a byproduct of strong obscuration in the Sgr B region. Blue: Continuum-subtracted soft X-ray line emission. The continuum subtraction removes the contribution from the dust-scattering haloes around bright point sources, which otherwise dominate the emission west of the southern chimney. }
\label{Molinari}
\end{figure*}
The left panel of Fig. \ref{Molinari} shows in red the soft X-ray emission map (1.5-2.6 keV), in green the S~{\sc xv} line emission (2.35-2.56~keV), and in blue the \meerkat\ map. The S~{\sc xv} line emission shows clear gradients in rough coincidence with the edges of the \meerkat\ bubbles. Because of the relatively high brightness and high energy emission of the S~{\sc xv} transition, this emission line provides us with an excellent tool to trace the hot plasma all the way to the Galactic plane. Indeed, it is significantly less affected by interstellar absorption than the soft X-ray emission map. 
The right panel of Fig. \ref{Molinari} shows the total and continuum-subtracted maps in the 1.5-2.6~keV band in red and blue, respectively. 
The continuum subtraction efficiently removes the emission from dust scattering haloes around bright sources (e.g., at: (l,b) = (359.98$^\circ$, 1.26$^\circ$); (359.56, -0.39); (359.30, -0.88); (359.08, -1.51); (359.12, -0.10); (0.67, 1.18); etc.) as well as nonthermal X-ray sources (e.g., pulsar wind nebulae), as is evident by comparing the red and blue maps. 
The green color shows the atomic hydrogen column density map of molecular material in a logarithmic scale from $1.2$ to $60\times10^{23}$ cm$^{-2}$ as observed by \herschel\ (Molinari et al. 2011). The highest concentration of molecular material occurs within a few tens of parsecs from the Galactic plane. The high column density of cold material is likely to significantly obscure the X-ray radiation toward the densest regions in the plane (i.e., the Sgr~B complex). 

A clear depression is observed in the X-ray emission at $b\sim\pm(0.2-0.3^\circ)$. We note that foreground absorbing clouds with column densities on the order of $N_H\sim0.4-1\times10^{23}$ cm$^{-2}$ are present at those locations (Molinari et al. 2011), which is sufficient to account for the observed depressions in X-ray emissivities in terms of increased X-ray absorption. \\ 

\subsection{The infrared (\wise) maps}
\label{secwise}

\begin{figure*}
\begin{center}
\vspace{-0.5cm}
\includegraphics[width=1.20\textwidth,angle=-90]{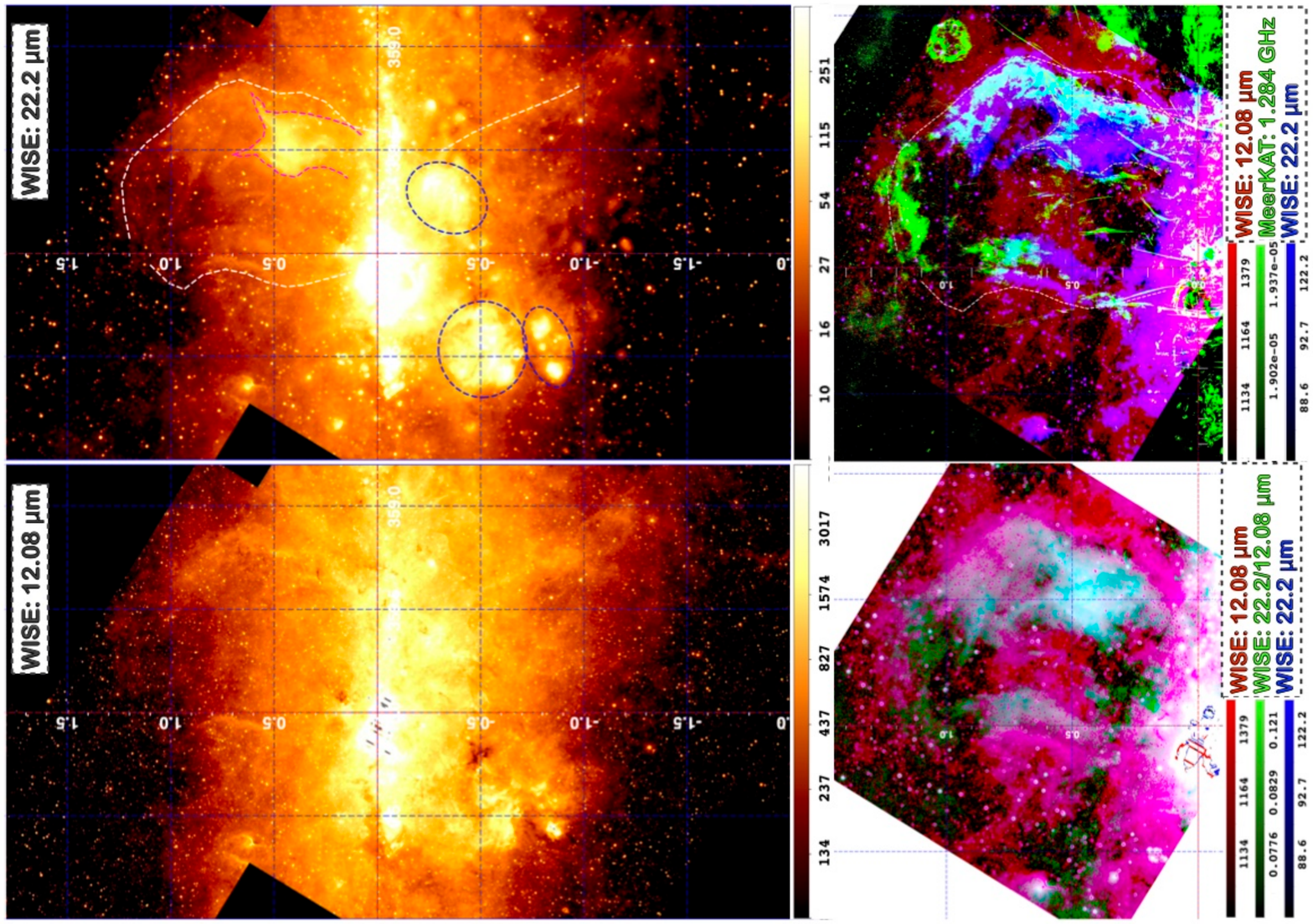}
\vspace{-0.7cm}
\end{center}
\caption{Mid-IR emission from the central degrees of the Milky Way. 
{\it (Top left)} Emission at 12.08~$\mu$m as observed by \wise. The highest concentration of warm dust is distributed along the Galactic plane. Spurs of material are clearly observed to emerge vertically from the plane. 
{\it (Top right)} Emission at 22.2~$\mu$m as observed by \wise. The most prominent spurs of matter are highlighted by white dashed lines. At northern latitudes both the western and eastern spurs are clearly associated with the northern chimney and delineate a lobe clearly associated with the observed radio emission. The magenta dashed line delineates the silhouette of \af. The detection of similar spurs in the southern hemisphere is complicated by the presence of well-known, bright foreground star-forming regions (highlighted by the blue ellipses). 
{\it (Bottom left)} Zoom-in toward the northern chimney. 
Red, green, and blue show the 12.08~$\mu$m, the ratio 22.2~$\mu$m/12.08~$\mu$m, and 22.2~$\mu$m emission, respectively. Warm dust surrounds the entire chimney. Also, the 22.2~$\mu$m emission peaks inside the location of the western shock and of the 12.08~$\mu$m emission spur. \af\ lies inside the edge of the warm gas and traces a $\sim90$~pc long molecular shock.   
{\it (Bottom right)} Same as previous but substituting the ratio map with the radio emission observed by \meerkat. The radio emission is very well correlated with the warm dust emission. In particular, the radio emission traces the perimeter of \af\ more tightly than the warm dust emission. }
\label{wise}
\end{figure*}

The \wise\ data were downloaded from the Infrared Science Archive\footnote{https://irsa.ipac.caltech.edu/applications/wise/} and subsequently mosaicked, adjusting the background to match in the overlapping regions\footnote{We stress that the constant background of each sky tile was adjusted arbitrarily to match in the overlapping regions and to facilitate the display of the various features. In particular, because of the arbitrary subtraction of constant backgrounds and because of the unknown contribution from extended foreground and background emission sources, the map made of the ratio of bands W4 (22.2 $\mu$m) to W3 (12.08 $\mu$m) is meant only to be indicative of the trends of dust temperature, and does not provide a quantitative measure of colour temperature.}, to obtain a full coverage of the chimneys. 
The top panels of Fig. \ref{wise} show the emission at 12.08 and 22.2~$\mu$m as observed by \wise. As for the dense, neutral material, 
the highest concentration of warm dust is distributed along the Galactic plane, however showing a considerably larger latitudinal extent. Indeed, spurs of warm dust (the most prominent of which are highlighted by white dashed lines) are clearly observed to emerge almost vertically from the plane. The blue dashed ellipses indicate the positions of well-known, bright, foreground star-forming regions. 

The bottom left panel of Fig. \ref{wise} displays an enlargement of the \wise\ map toward the northern chimney, where in red and blue are shown the 12.08 and 22.2~$\mu$m emission, respectively. The green colors show the ratio of infrared color defined as the 22.2~$\mu$m map divided by the 12.8~$\mu$m one. The bottom right panel shows in red and blue the \wise\ maps (as in the left panel) and in green the radio emission as observed by \meerkat. The white dashed lines display the location of the spurs delineated on the basis of the \wise\ maps. 

\section{Discussion}

We start the discussion by emphasizing two points. \\

{\bf First point:} The overall degree of symmetry of the radio and X-ray emission (see Fig. \ref{xmmMK}) around the GC suggests that the chimneys are a single coherent feature located at the GC. Additionally, Fig. \ref{all} demonstrates that the X-ray, radio and infrared emission are deeply connected. Indeed, Fig. \ref{all} shows that they form coherent features, extending for hundreds of parsecs, that can be followed both in the X-rays, radio and the infrared bands. This demonstrates that the main players traced in each band, which are the hot plasma in X-rays, the warm dust in infrared and the shocks in radio, are interacting and deeply affecting each other. This strengthens the idea that they are all byproducts of a single energetic phenomenon. \\

{\bf Second point:} We note from the radio image (Fig. 3) that the surface density of well-defined nonthermal filaments is highest within $\sim$0.5 degrees of the Galactic plane and drops rather abruptly with latitude beyond that. In some cases, the narrow non-thermal filaments become increasingly diffuse at higher latitudes, especially above and below the Radio Arc.  In addition, at northern latitudes, the magnetic field lines delineated by the nonthermal filaments appear to diverge with increasing latitude. We conclude from these observations that the magnetic flux decreases with increasing Galactic latitude (see also Morris 2006b; 2015). Such a decrease would affect how the GC outflow is manifested as a function of latitude.  At low latitudes, the shock occurring where the outflow impacts the surrounding ISM would encounter a relatively stronger field, and it would encounter it at a steep angle, so that the shock is likely to be a C-type shock in which the shock energy is distributed broadly over a relatively large region (c.f., Draine 2011), and without a velocity jump large enough to ionize the gas passing through the shock\footnote{Indeed the shock compression of the vertical field by a C-type shock at the location of the Radio Arc might play a role in generating the bundle of nonthermal filaments constituting the Arc.}.  At high latitudes, however, with a weaker field and a more oblique shock, in the presence of a predominantly vertical field, the velocity jump could be sufficient to ionize the gas in the shock. The diffuse radio emission that we associate with the shock induced by the GC outflow as it impacts the surrounding medium would therefore appear most prominent at the higher Galactic latitudes, as is observed. \\

The following subsections will discuss the various components of the chimneys. Section~\ref{secArc} considers the multi-wavelength emission from the Arc super-bubble, which might be instructive for a deeper understanding of the chimneys. 
In \S \ref{foreground} we debate the proposed foreground location of part of the northern chimney.
In \S~\ref{NW} we examine the northwestern edge of the chimneys, reporting evidence for a multiphase (i.e., hot and cold-molecular) and multi-epoch outflow that produces strong shocks at the edges of the chimneys. In Section~\ref{cap} we talk about the northern "cap." In \S~\ref{east}, we discuss the eastern edge of the chimneys, highlighting the differences compared with the western edge, stressing the importance of the GC magnetic field and introducing the concept of a "magnetic wall." Section \ref{protrusion} discusses possible origins of the X-ray protrusion. Section~\ref{south} considers the southern chimney. 
Section~\ref{highb} shows that the continuity of hot plasma emission all the way to the higher latitudes, where the Fermi bubbles begin, provides evidence that the chimneys are the multi-epoch outflows that feed the Fermi bubbles with energetic particles.
Finally, in \S \ref{fila} we ask whether the nonthermal radio filaments might be associated with the GC outflow. 

\subsection{The Arc super-bubble as a template to understand the chimneys}
\label{secArc}

\begin{figure}
\hspace{-0.4cm}
\includegraphics[width=0.59\textwidth,angle=0]{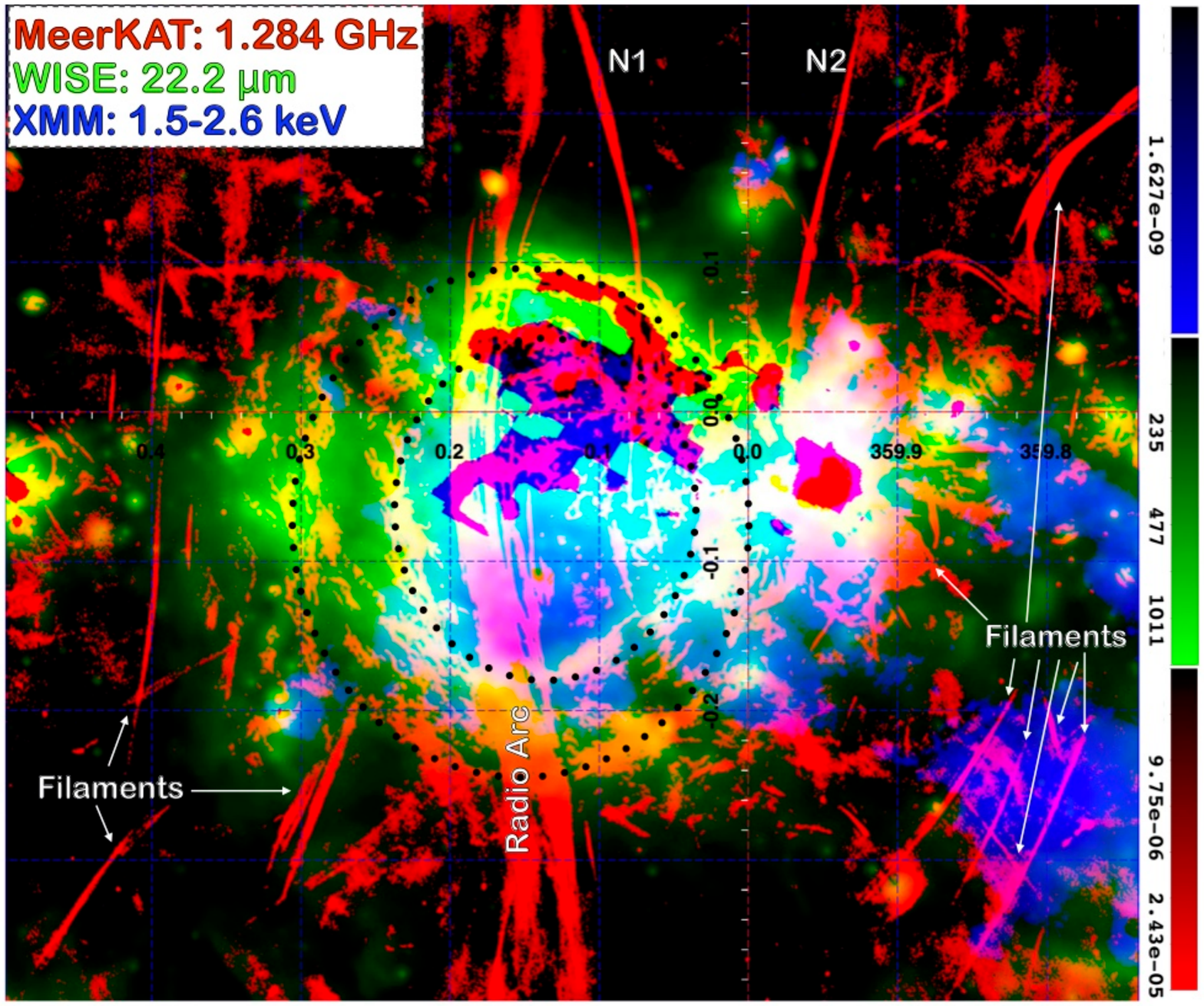}
\caption{Multiwavelength view of the Arc super-bubble. Red colors show the \meerkat\ map. Green colors show \wise\ 22.2~$\mu$m emission and the blue colors show the soft X-ray emission map (1.5-2.6 keV) of the region around the Arc super-bubble. The radio and mid-infrared emission are edge-brightened and trace the edges of the Arc super-bubble, while the X-ray emission appears to fill the central volume. 
Despite being edge-brightened, several bright stripes and arcs of radio emission, presumably tracing shocks, are projected toward the interior of the Arc super-bubble, as they are for the chimneys (some are highlighted by the two black dotted ellipses). The Radio Arc as well as many narrow, nonthermal filaments (some of which are highlighted by the white arrows) are clearly visible (see Fig. \ref{FC} for a definition of the Radio Arc and most prominent nonthermal filaments). Saturation and bright source removal in the \wise\ and \xmm\ maps produce some visible artifacts. }
\label{Arc}
\end{figure}

Figure \ref{Arc} shows a radio (red), mid-infrared (green) and X-ray (blue) overlay centered on the Arc super-bubble. The super-bubble appears brightly in all of these bands. Figure \ref{Arc} fully supports the paradigm that the Arc super-bubble is filled with hot plasma occupying an X-ray bright "cavity," surrounded by warm, shock-heated dust (Egan et al. 1998; Levine et al. 1999; Rodriguez-Fernandez et al. 2001; Price et al. 2001; Sofue 2003; Simpson et al. 2007; Ponti et al. 2015). Indeed, bright mid-infrared emission is observed along the entire limb-brightened perimeter of the Arc super-bubble. The radio emission runs along the circular ridge of mid-IR emission and encloses the  X-ray emission from the interior of the super-bubble. 
In addition, patches of bright radio emission can be observed also projected toward the interior of the Arc super-bubble (see the dotted ellipses highlighting this emission in Fig. \ref{Arc}), corroborating the idea that the radio emission is primarily tracing ionization fronts and also shocks at the interface of the super-bubble with the ISM. 
A detailed description of such an array of concentric radio shells (observed in the historical VLA 20cm data; Yusef-Zadeh \& Morris 1987a,b) can be found in Sofue (2003). That work attributed the creation of the shells to recent ($\sim10^6$~yr) starbursts, the most likely candidate being  the supernovae and stellar winds from the Quintuplet cluster (Egan et al. 1998; Sofue 2003; Ponti et al. 2015).

Therefore, the Arc super-bubble represents a textbook example of a powerful outflow within the GC environment that is currently still contained within the disk of the Milky Way (Fig. \ref{Arc}). Although the more powerful chimneys apparently succeeded in breaking through the Galactic density gradient and in overcoming the Galactic potential, thereby opening a channel to the halo, we expect that their multiwavelength emission maintains some degree of resemblance to the Arc super-bubble.
Indeed, as with the Arc super-bubble, within the chimneys we also observe intense radio emission tracing shocks occurring primarily at their edges and X-ray emission located primarily inside the chimneys as well as warm dust emission at various locations along the borders of the chimneys. 

\subsection{On the recent suggestion that the northwestern edge of the chimney is a foreground feature}
\label{foreground}

The observation of the silhouette of the northwestern and northern portion of perimeter of the northern chimney against low-frequency radio emission (LaRosa et al. 2005; Brogan et al. 2003; Hurley-Walker et al. 2019; Tsuboi et al. 2020) clearly implies that the northern chimney is located in front of most of the diffuse radio continuum, which is produced primarily within the inner few hundred parsecs. 
It is currently highly debated whether the northwestern chimney is located along the Galactic disk at a few kiloparsecs from the Sun (Nagoshi et al. 2019; Tsuboi et al. 2020; Wang 2020) or whether it is instead placed just in front of the GC. 
The overall morphological symmetry of the radio and X-ray chimneys strongly suggests that at least some portion of the northwestern edge is located at the GC. Additionally, the association of the northwestern edge of the chimney with the \af\ feature (see \S \ref{NW}), which is characterized by a very high positive velocity, indicates that both \af\ and the northwestern edge of the chimney are located at the GC. 
We note that, on the sky plane, the chimneys have a small overall tilt of about 7$^{\circ}$ with respect to the vertical to the Galactic plane, perhaps as a result of local pressure gradients, cloud placements, and initial injection directions.  In any case, this inclination of the chimneys raises the possibility that the northern chimney is also inclined toward us to some extent.  If the tilt is such that the northern chimney is inclined toward us by a few tens of degrees, then the majority of the northern chimney would be located in front of the bulk of the GC radio continuum, which could account for its appearing in absorption in low-frequency radio maps, while its southern counterpart, if the chimneys are indeed colinear, would cast no shadow on those images, as observed (more detailed discussion on this topic is presented in \S \ref{appendix}). 
Therefore, hereinafter, we will assume that the northwestern chimney is part of the chimneys that is rooted at the GC. 
However, we warn the reader that projection effects might be important and that a portion of the features here assumed to be located at the GC might be foreground features unrelated with the Galactic outflow. 

\subsection{Northwestern edge}
\label{NW}

\subsubsection{Entrained molecular outflow producing a $\sim0.1$~kpc long shock onto the ISM}  
Figure \ref{xmmMK} shows that the western edge of the chimneys is very well defined, both in X-rays (see thick dashed line in Fig. \ref{xmmMK}; Ponti et al. 2019) and radio. Indeed, the radio emission runs parallel to the X-ray edge along its full extent of $\sim350$~pc, that is, all the way from the northern part of the northern chimney to the southerly tip of the southern chimney. However, the peak of the radio emission on the western side of the chimneys is displaced toward more negative longitudes by $\sim0.2^{\circ}$ ($\sim30$~pc) than the western edge of the X-ray emission. 
Additionally, Figure \ref{wise} shows that warm dust surrounds the entire northern chimney. Indeed, both the 12.08 $\mu$m and the 22.2 $\mu$m maps show a spur of material (IR1, see Fig. \ref{FC}) emerging from the Galactic plane (at the location of the Sgr~C molecular complex) and reaching a cap of material observed at high Galactic latitudes around $b\sim1.1-1.4^\circ$ (Fig. \ref{xmmMK}, \ref{FC}, \ref{wise} and \ref{all}). Again we note that, while the X-ray, radio and warm dust distributions run parallel to each other, they are significantly displaced in longitude (by $\sim0.1-0.3^\circ$). This is discussed further below.

\subsubsection{AFGL~5376: The brightest part of a molecular shock}

Almost three decades ago, Uchida et al. (1994) studied the molecular line emission around this region and discovered two high-velocity components of molecular material, defining a vertical rift coinciding with the vertical ridge of the mid-IR source, \af.
The large velocity separation of the two molecular components — 65 km s$^{-1}$ — led Uchida et al. (1994) to suggest that the strong IR emission from \af\ results from the energy deposited in a 90-pc long shock where the two molecular components meet (Fig. \ref{shocks}). Furthermore, the large positive velocities of both components are best explained by expansion motions away from the GC, so we presume that they are participating in the GC outflow\footnote{The line of sight velocity of \af\ is consistent with the one expected if such a cloud were located along the innermost non-self-intersecting X1 orbit induced by the Galactic bar of stars (see e.g., Binney et al. 1991). If so, the velocity of \af\ would not be due to expansion, but rather to the streaming motions of gas along the bar, characteristic of the X1 orbits. However, such a scenario leaves two things completely unexplained: 1) the fact that \af\ is rather far out of the Galactic plane, and 2) the fact that \af\ is the site of such a strong internal shock. For these reasons, we believe that \af\ participates in the outflow from the GC region. }. The authors hypothesize the presence of two shocks, represented by the thick-solid blue lines in Fig. \ref{FC}. The western blue arc would represent the leading edge of a strong ionizing shock with a westward velocity component, delineating the location where the outflow encounters the ISM (see also Fig. \ref{shocks}). A shock with $v\ge65$ km s$^{-1}$ could dissociate CO and H$_2$ molecules and ionize hydrogen, inducing free-free emission. The eastern blue line, which coincides with the IR ridge, was suggested to represent a dissociation front, associated with the reverse shock, where the fast outflowing material encounters the more slowly moving post-shock material preceding it (Uchida et al. 1994). 

\subsubsection{Corroborating the shock-interpretation with fresh data}

\begin{figure}
\hspace{-0.45cm}
\includegraphics[width=0.55\textwidth,angle=0]{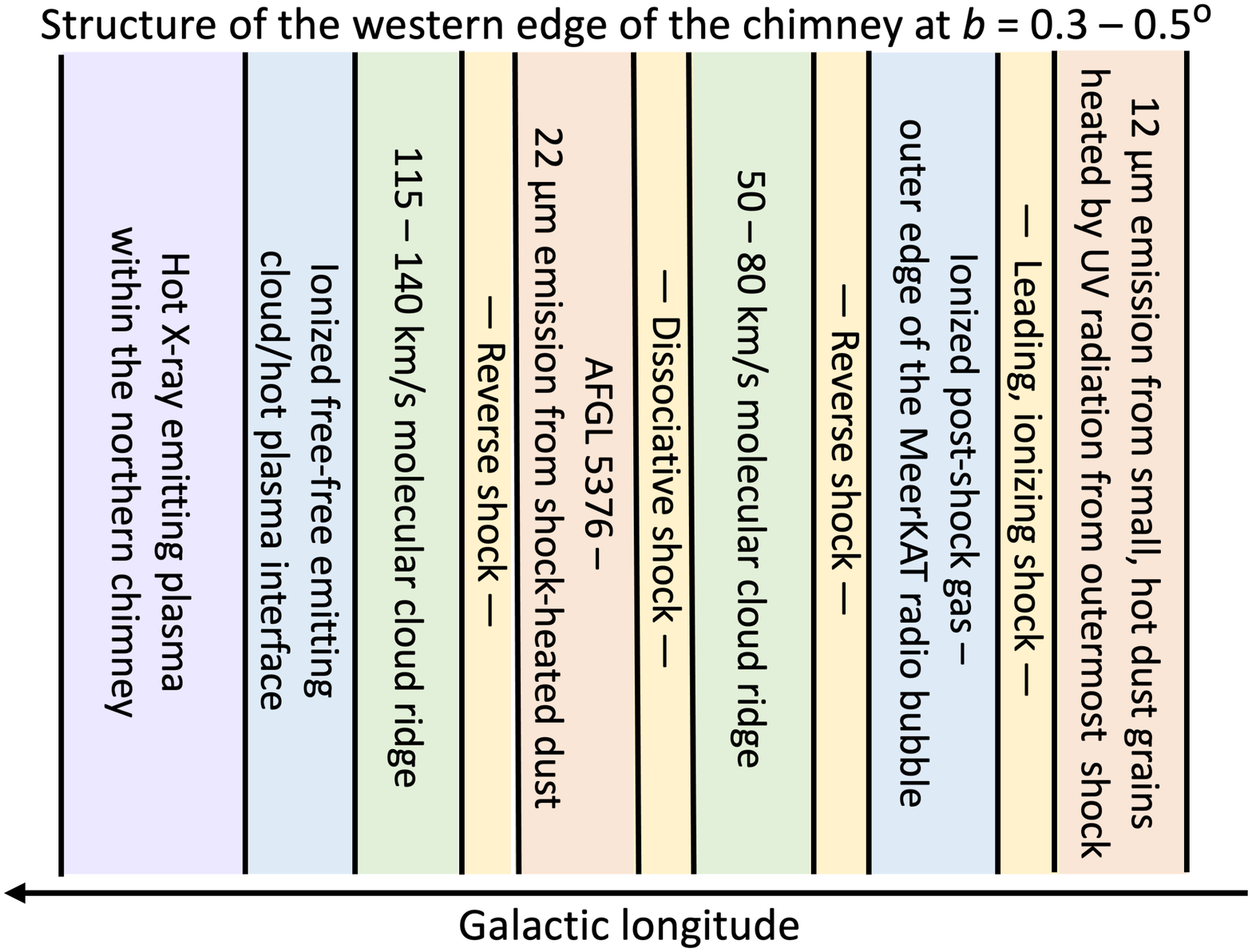}
\caption{Schematic representation of the structure of the northwestern edge of the chimney. }
\label{shocks}
\end{figure}

The data presented here corroborate this interpretation. Indeed, the \wise\ versus \meerkat\ overlay shows intense radio emission downstream of the strong shock, on the western side of \af\ (Figs. \ref{wise} and \ref{all}). Additionally, the current data reveal that the entire perimeter of the \af\ feature is bright in radio emission, supporting the idea of a structured interaction between the outflow and the ISM, possibly through a secondary (or "reverse") shock between \af\ and the surrounding ISM (see Fig. \ref{shocks}). 
Such a reverse shock is consistent with being a dissociative shock. Indeed, the lack of radio emission running down the infrared ridge at the center of \af, disfavor it being an ionizing shock. 
Additionally, the radio emission on the eastern perimeter of \af\ is at the interface between the X-ray emitting plasma and the molecular cloud, so that ionization could be induced by particle impact (e.g., conduction). 

Figure \ref{all} reveals that almost no X-ray emission is observed between the strong and reverse shock, in agreement with the idea that the shocked molecular outflow, visible as warm dust, might form a structured shock with the ISM, with the low-density, shock-heated plasma radiating in the X-ray band located in the internal post-shock region, inside of longitude $\sim359.6^\circ$ (Fig. \ref{all}; \ref{shocks}). Finally, the \wise\ maps (Fig. \ref{wise}) clearly show the presence of warm dust running along the edge of the strong shock proposed by Uchida et al. (1994). 
We point out that it remains unclear whether \af\ impacts the ambient ISM "head on" or whether there is considerable shear parallel to the shock. 

The extraordinary result revealed by the superposition of the \xmm, \meerkat\ and \wise\ maps is the discovery that such an association of X-ray, radio and mid-infrared emission is not confined to the small region next to \af, but it continues to latitudes of $b\sim1.2^\circ$ and all the way to the cap (Fig. \ref{all}). This suggests the presence of a shock, of heated dust and swept up material accumulating at this edge of the chimneys. 

The \meerkat\ map shows that the radio emission traces the presumed location of the strong shock defining the northwestern chimney (Figs. \ref{xmmMK} and \ref{all}). The ratio of the 22.2 $\mu$m over the 12.08 $\mu$m emission shows that the 22.2~$\mu$m emission along most of the northwestern edge, traces the radio emission at the location of the shock, as expected for shock heated dust (Fig. \ref{wise} and \ref{all}). Furthermore, the 22.2 $\mu$m emission extends for many (up to tens of) parsecs behind the shock itself, indicating a strong interaction between the GC outflow and the mechanism heating the dust. 

\subsubsection{Indications of stochastic heating of pre-shock dust grains}

Figure \ref{wise} shows that the peak of the 12.8 $\mu$m emission along most of the northwestern edge occurs either at the presumed location (or even outside) of the strong shock defining the northwestern edge of the chimney (Fig. \ref{wise}). 
At first sight, this appears surprising because the 12.08~$\mu$m emission typically traces warmer dust than the 22.2~$\mu$m one, leading us to the paradox of having warmer dust ahead of the shock than behind it. 
This might be understood if the strong, ionizing shock acts as an efficient source of the photons that stochastically heat the small dust grains, producing enhanced 12.08~$\mu$m emission, even ahead of the shock front, despite the low temperature of the larger dust grains in the pre-shock region. After the shock, all the dust gets significantly heated, not just the small grains, therefore enhancing the 22.2~$\mu$m emission in addition to the emission at 12.08~$\mu$m. We point out that the mean free path of the diffuse Lyman continuum is likely too short ($\sim0.01$~pc for a density of 10~cm$^{-3}$; Draine 2011) to contribute to such an effect\footnote{Even shorter mean free paths are associated with Lyman $\alpha$ photons. }. Therefore, we speculate that if the Ly $\alpha$ emission is produced in a region with high velocities, then the Lyman $\alpha$ line might be sufficiently Doppler shifted to be out of the local line profile, thus reducing the optical depth considerably. 
In support of such a scenario, the overlay of the \wise\ versus \meerkat\ maps (bottom right panel of Fig. \ref{wise}) shows, along the full extent of the northwestern spur (IR1) and cap, intense radio emission (tracing the shock front) peaking in the zone where the 12.8~$\mu$m start dominating over the 22.2 $\mu$m  emission, with increasing distance from the center.  

\subsubsection{\af: Evidence for a multiphase outflow}

For a mass of \af\ of $M_{\af}\sim5\times10^4$~M$_\odot$, an estimated volume of $V_{\af}\sim18^2\times7$~pc$^3$ and a relative velocity to the pre-shock gas of $v_{\af}\sim65$~km~s$^{-1}$, we obtain a density of $\sim10^3$~cm$^{-3}$ and a ram pressure of $p_{ram}\sim6\times10^{-8}$ dy cm$^{-2}$, which corresponds to $\sim40$~keV cm$^{-3}$. Such pressure is about two orders of magnitude larger than the thermal pressure of the observed hot plasma within the chimneys, which has been estimated to be $p_{hot}\sim0.1-0.2$ keV cm$^{-3}$ (Ponti et al. 2019). Assuming that the thermal pressure of the hot plasma is accelerating \af, we estimate that it would require $>10^7$ years to reach a speed of 65~km~s$^{-1}$ or more. 
Even if the hot plasma has a subsonic bulk motion as high as $v\sim500$~km~s$^{-1}$, the ram pressure then results to be on the order of $\sim0.5$ keV cm$^{-3}$, therefore still requiring a long time to accelerate \af. 
Therefore, this appears to disfavour models in which \af\ has been accelerated by the hot plasma that we currently observe within the chimneys, suggesting that an alternative agent might be needed to account for the outflow speed of the molecular cloud.   

\subsubsection{Origin of the \af\ feature}

We note that the origin of \af\ might be connected with that of the recently discovered high-velocity HI clouds located at the base of the Galactic bulge (McClure-Griffith et al 2013; Di Teodoro et al. 2018; 2020; Lockman et al. 2020). Indeed, recent large-scale HI surveys of the Galactic bulge led to the discovery of a large population ($\sim200$) of anomalous high-velocity HI clouds extending up to $\sim3$~kpc north and south of the Galactic plane, in a bi-conical configuration, with a neutral cloud lifetime of $\sim4-10$~Myr and with kinematics consistent with an outflow from the GC. The neutral cloud component of the outflow is characterized to have an opening angle of $>140^\circ$, a maximum outflow velocity of $v_{out}\sim330$~km~s$^{-1}$, a mass outflow rate of $\sim0.1$~M$_\odot$~yr$^{-1}$, and a kinetic luminosity of the outflow of $L_k>3\times10^{40}$ erg s$^{-1}$ over the past 10~Myr (McClure-Griffith et al 2013; Di Teodoro et al. 2018; 2020; Lockman et al. 2020). For densities of such HI clouds in the range $n_H\sim1-20$~cm$^{-3}$ and local standard of rest velocities of the clouds in the range $v_{LSR}\sim100-360$~km~s$^{-1}$, the driving wind should be imposing a ram pressure of $\sim0.1-30$~keV~cm$^{-3}$ to accelerate such clouds.  The upper bound on the required pressure range is one-to-two orders of magnitude larger than the thermal pressure of the hot plasma currently observed within the chimneys and similar to the ram pressure needed to accelerate \af. We suggest that the process that accelerated the HI clouds distributed throughout a biconical volume at the base of the \fermi\ bubbles has also accelerated some massive clouds in directions closer to the Galactic plane, and is thus responsible for the dynamics of \af. 

\subsubsection{Further evidence for a multiphase outflow}

Other high-velocity molecular features are observed toward the chimneys, further supporting the multiphase nature of the outflow (Hsieh et al. 2015; 2016). The polar arc represents one such example. The polar arc is an extra-planar molecular cloud, located $\sim30$~pc above \sgras, with a high radial velocity ($v>+100$~km~s$^{-1}$) and a positive velocity gradient perpendicular to the Galactic plane, suggesting an accelerating and expanding motion off the plane (Bally et al. 1988; Hsieh et al. 2015; 2016). This cloud appears connected with an extra-planar hourglass-shaped feature having an extent of $\sim13$~pc perpendicular to the plane and centered on the central parsec of the Milky Way, and a dynamical timescale of $\sim3\times10^5$~yr (Hsieh et al. 2016). The hourglass feature runs along the edges of Sgr~A's bipolar lobes containing hot plasma (see the remarkable agreement represented in Fig. \ref{FC} by the hourglass feature in blue and the edges of Sgr~A's bipolar lobes in orange; Morris et al. 2003; Ponti et al. 2015). Indeed, the observed molecular components might represent the entrained molecular gas within the hot plasma of Sgr~A's bipolar lobes (Hsieh et al. 2016; Ponti et al. 2015; 2019). The morphology and kinematics of these molecular clouds within tens of parsecs of the GC are reminiscent of kpc-scale molecular outflows in nearby starburst galaxies (Garcia-Burillo et al. 2001; Walter et al. 2002; Bollato et al. 2013). Indeed, they are consistent with an origin in the Galactic plane from which the clouds have been lifted. As a matter of fact, they have been suggested to have orbital paths altered either by an explosion $\sim10^5$~yr ago or by outflows from the central parsec of the Milky Way (Hsieh et al. 2015; 2016). 

A recent extensive survey of H$_3^+$ absorption toward bright stars within the central molecular zone demonstrated the presence of an outflow of warm ($T\sim200$~K) diffuse ($n\sim50$~cm$^{-3}$) gas (Oka et al. 2020). The diffuse, warm gas is observed to expand with radial velocities of $\sim150$~km~s$^{-1}$ and to be as extended as $\sim150$~pc from \sgras\ (Oka et al. 2020). The energy, momentum and timescale required to create such an outflow have been estimated to be $\sim5\times10^{53}$~erg, $5\times10^8$~M$_\odot$~km~s$^{-1}$ and $\sim(0.5-1)\times10^6$~yr (Oka et al. 2020).  This revives the idea of either an expanding ring (Kaifu et al. 1972; Scoville 1972) or an expanding bi-polar vertical cylinder with total length as great as $\sim170$~pc (Sofue 2017). 

\subsubsection{Evidence for a multi-epoch outflow}

Despite the large uncertainties (e.g., line of sight distance, 3-d velocity, launching point, etc.), it appears that the molecular clouds associated with \af\ were first launched onto their relatively high-latitude trajectories a few $10^6$~yr ago. We note that this timescale is comparable to the one associated with the outflow of warm diffuse gas, while it is about 10 times longer than that of the hourglass-shaped feature. Additionally, it is $\sim10-100$ times longer than the sound crossing time of the chimneys ($t_s\sim3\times10^5$~yr) and of Sgr~A's bipolar lobes ($\sim3\times10^4$~yr) and more than ten times longer than the relatively recently formed recombining plasma, with estimated age less than $\sim10^5$~yr (Nakashima et al. 2013). We 
conclude that these greatly different timescales indicate that the chimneys were shaped by events occurring at widely different times, and therefore they are the product of multi-epoch events. 

\subsection{The cap}
\label{cap}

The superposition of the radio, mid-infrared and X-ray images reveals the presence of a cap above the northern chimney (see Figs. \ref{xmmMK}, \ref{FC}, \ref{wise} and \ref{all}). This is clear evidence that the GC outflow to the north has swept up material, or is being at least partly impeded by moderately dense material there. 

We speculate that the different latitudinal extent and longitudinal width of the northern and southern chimneys can be attributed to a greater initial ISM mass in the northern volume, leading to more mass swept up to create the "northern cap" and to a broader longitudinal extent because the enhanced confinement to the north causes a greater pressure that leads to a greater expansion in the longitudinal direction.  This is also consistent with the fact that the \af\ cloud and the polar arc cloud are located to the north, but no comparable clouds have been seen near or in the volume occupied by the southern chimney.

We note that the X-ray maps show a clear gradient of decreasing hot plasma emission (clearer in the S~{\rm xv} map; Fig. \ref{Molinari}) at the location of the cap. However, intense hot plasma emission is also observed beyond the cap (Fig. \ref{xmmMK}, \ref{Molinari}, \ref{all}). 
In particular, the surface brightness, temperature, density and pressure of the hot plasma are consistent with a rather smooth transition across the cap, suggesting that the outflow associated with the last event of a quasi-continuous series of intermittent energy releases is only partially impeded by the material in the cap.  

\subsection{Northeastern edge}
\label{east}

\subsubsection{Differences between the eastern and western edges}

Because of the absence of a structured shock similar to that seen at the northwestern boundary, the eastern edge of the chimneys is less well defined (Fig. \ref{xmmMK}). The \meerkat\ map indicates an association of the eastern edge of the chimneys with the Radio Arc, however it also reveals that the morphology of the eastern edge, within $\sim0.5^\circ$ of the plane, is dominated by bundles of filaments, therefore very different from the "fuzzy" appearance of the northwestern edge and cap. We also note that the filamentary appearance of the eastern edge becomes fuzzier at higher latitudes. This could be understood in terms of a magnetic field that dominates the pressure at the eastern edge of the chimneys, but starts to diverge beyond $|b|\sim0.5^\circ$, and therefore become less confining of the synchrotron emitting particles that occupy the filaments.

Additionally, the radiative mechanism (synchrotron vs. free-free emission) and degree of polarization are two of the major differences between the eastern and western edges of the northern radio bubble, which implies that the northeastern edge is undergoing very different physical processes compared with the western edge (Fig. \ref{xmmMK}, \ref{FC}, \ref{wise}, \ref{all}; see also Reich et al. 1987; Haynes et al. 1992). 

Finally, the morphology of the hot plasma emission along the northeastern edge represents another major difference compared with the northwestern edge. Very intense X-ray emission is observed all the way from the interior of the chimney to and beyond the location of the Radio Arc. The X-ray emission at that location is primarily thermal (with temperatures of $\sim0.7-1$~keV; Ponti et al. 2019), therefore we exclude a major contribution due to synchrotron emission associated with the nonthermal filaments, although X-ray counterparts to the filaments do make a minor contribution (Wang et al. 2002, 2020 preprint, Zhang et al. 2014; Ponti et al. 2015; Mori et al. 2015).

\subsubsection{Hints of a dominant vertical magnetic field}

We note that on the eastern edge of the chimney, the radio map offers many manifestations of a pressure-dominant magnetic field, including: i) exceptionally long nonthermal radio filaments such as the N1, N2 and XMM~J0.173‐0.413 filaments, running north and south of the plane; ii) the Radio Arc; and iii) the highly polarized radio plumes running both north and south of the Galactic plane. All of these features underline the presence of a strong (likely dominating the ambient pressure) magnetic field with a strength that some have estimated to be as high as a few mG (Seiradakis et al. 1985; Morris \& Yusef-Zadeh 1985; Tsuboi et al. 1986; Yusef-Zadeh \& Morris 1988; Lang et al. 1999; Blanton 2008; Magilli et al. 2019). 

Also, the mid-infrared spur (IR2, see Figs. \ref{wise} and \ref{FC}) can be best understood in the framework of a high magnetization. We note that, close to the Galactic plane, IR2 runs parallel to the N1 filament up to $b\sim0.4-0.5^\circ$, which is radiating synchrotron emission due to relativistic electrons in a highly ordered vertical magnetic field. Between $b\sim0.4$ and $b\sim0.7^\circ$, the mid-infrared spur is defined by the Double Helix Nebula (DHN; top left panel of Fig. \ref{wise}; Morris et al. 2006; Tsuboi et al. 2010). The radio continuum emission in the general direction of the DHN is observed to be highly polarized, indicating a highly ordered magnetic field with synchrotron-emitting relativistic electrons (Tsuboi et al. 2010). Molecular line surveys of this region have revealed two molecular components at 0 and $-35$~km~s$^{-1}$ associated with the DHN, with no clear evidence of shocks and with turbulent line broadening of $\sim3-5$ km s$^{-1}$. The mass of the two molecular components is estimated at $\sim3.3\times10^4$ and $\sim0.8\times10^4$~M$_\odot$, respectively (Enokiya et al. 2014; Tori et al. 2014). These authors, following Morris et al. (2006), propose that the warm dust has been forced into such a double helix configuration by a strong magnetic field. 

Torii et al. (2014) inferred densities of the DHN of $n_e\sim0.7-2\times10^3$~cm$^{-3}$. Assuming such densities and that the matter within the DHN is shaped into such a configuration by a turbulent pressure that is also the source of the line broadening, we compute the turbulent pressure to be in the range of $\sim0.1-0.5$~keV~cm$^{-3}$. Such a pressure would be in balance with the magnetic pressure for a magnetic field with a strength of $0.1$~mG. Therefore, the vertical magnetic field with a strength of $\sim1$~mG (whose presence is corroborated by the presence of the nearby Radio Arc, filaments and polarized plumes) would, indeed, be able to dominate the dynamics of the DHN. 

\subsubsection{A magnetic wall?}

We point out that if, as suggested, on the eastern side of the chimneys the magnetic field has a strength as high as $\sim1$~mG and a uniform vertical configuration, then it cannot be neglected when considering the dynamics of the hot plasma\footnote{Such a scenario has been already considered by several authors. Sofue (2020a), for example, considers a magnetic cylinder as the origin of the chimneys and nonthermal filaments. He hypothesizes that magnetohydrodynamic compression waves ejected from the nucleus might be reflected and guided through the magnetic field, therefore appearing as the nonthermal filaments (when viewed tangentially) and produce feedback loops between accretion events and magnetic outflows Sofue (2020b).}. 
Indeed, the longitudinal expansion of any outflow with a pressure lower than $\sim100$~keV cm$^{-3}$ (corresponding to a magnetic field strength of 1-2~mG) would encounter the resistance of the magnetic field, while it would allow the flow to continue almost unperturbed perpendicular to the Galactic plane, therefore with the net effect of collimating the outflow. We also note that, adding to the collimation by the magnetic field, the vertical density gradient in the Galactic plane would contribute to the collimation. In this scenario, an outflow would be currently flowing through the chimneys. 
Alternatively, the hot plasma emission would be the relic of a past major outburst at the GC, now in hydrostatic equilibrium within the gravitational potential of the Milky Way and still radiating because of the very long cooling time ($2\times10^7$~yr; Ponti et al. 2019). In such a scenario, it seems plausible that the hot plasma was generated and then remained at the edge of the outflow. 

Despite its pivotal importance, this scenario fails to explain what is sustaining the magnetic field. Several authors have connected the creation of the strong (in the mG range), vertical GC magnetic field to the accretion of plasma through the Galactic disk over the entire life of the Milky Way (Sofue et al. 1987; 2010; Sofue \& Fujimoto 1987; Howard \& Kulsrud 1997; Chandran et al. 2000). In such scenarios, the toroidal component of the field would be amplified by differential rotation, and toward the GC, radial compression would amplify the vertical field, creating a nearly vertical magnetic field because of inefficient ambipolar diffusion (Chandran et al. 2000). In particular, Sofue et al. (2010) showed that the winding of the primordial magnetic field can evolve into composite configurations, comprising bisymmetric spiral, axisymmetric spiral, plane-reversed spiral, and/or sing fields in the disk, and vertical fields in the center, similar to what is observed in nearby spiral galaxies. 
We speculate that, if the magnetic field is mass-loaded, then its more relevant role on the eastern side of the GC might be associated with the larger amount of mass on the Galactic plane at positive longitudes (Fig. \ref{Molinari}; Bally et al. 1988; Tsuboi et al. 1999; Molinari et al. 2011; Jones et al. 2012). Additionally, the effect of the magnetic field would appear more evident than on the northwestern side due to the fact that the ISM appears less dense above the plane on that side. 

\subsection{Protrusion}
\label{protrusion}

The X-ray map shows bright X-ray emission at $(l,b) \sim (0.4,0.4^\circ)$, which appears as a "protrusion" from the northern lobe (Fig. \ref{xmmMK} and \ref{all}; Ponti et al. 2019). The origin of such emission is unclear. 

The physical properties of the hot plasma (e.g., temperature, density, energetics) within the protrusion could be consistent with an interloper supernova remnant (SNR). To validate this hypothesis, we note that the \xmm\ and \meerkat\ maps show at least four SNRs at Galactic latitudes comparable to or higher ($|b|>0.3^\circ$) than that of the protrusion, therefore confirming the high projected density of SNRs in this region. 
Additionally, we note that, along the line of sight toward the Radio Arc a crescent-shaped radio feature is observed at the western border of the protrusion, at (l,b) = (0.07$^\circ$, 0.18$^\circ$), which is consistent with being a radio shell (Figs. \ref{MeerKAT}, \ref{xmmMK} and \ref{all}). Furthermore, radio emission is observed around most of the remaining portions of the protrusion, although it is weak (compare Figs. \ref{xmmMK} and \ref{MeerKAT}), and not definitively associated with the protrusion. Therefore, the lack of a clear radio counterpart in either \meerkat\ or previous radio surveys leaves the association of the protrusion with an SNR unclear (Haynes et al. 1992; LaRosa et al. 2000; Heywood et al. 2019). 

Figure \ref{Molinari} shows that the S~{\rm xv} emission (which is less affected by absorption than the soft X-ray emission map), extends the emission of the protrusion all the way to the Galactic plane at the location of the Sgr B1 and Sgr B2 molecular complexes. Indeed, the Sgr~B molecular complex represents the region of the Milky Way with the highest specific star formation rate (SFR$_{B2}\sim0.04$~M$_\odot$~yr$^{-1}$; Armillotta et al. 2019). 
Therefore, based on these data, we do not exclude the possibility that the protrusion might represent either a super-bubble or the early phases of the formation of a Galactic fountain, which could contribute to energizing and excavating the chimneys. 
Forthcoming deep radio maps of the protrusion hold the key to understanding the origin of such a peculiar hot plasma feature.  

\subsection{Southern chimney}
\label{south}

The study of the southern chimney is complicated by the presence of well-known bright foreground star forming regions down to latitudes of 
$\sim0.5^\circ$ and $\sim0.7-0.9^\circ$ (highlighted by the blue ellipses; Fig. \ref{FC}). Indeed, ambiguity remains regarding how much of the observed radio, X-ray, and mid-infrared emission is associated with foreground features. 

The western side of the southern chimney appears as an impressive extension of the northwestern edge (Figs. \ref{all} and \ref{xmmMK}). Indeed a single straight line well represents the border between the hot plasma distribution and the radio emission. This morphological correspondence suggests a direct link between the two in the form of a structured shock such as its northern counterpart, although not as clearly defined. 
The lack of an evident mid-infrared spur of warm dust and of shocked molecular clouds might be the consequence of its much shorter cooling time as well as of smaller ISM densities present at these negative latitudes (c.f., discussion in \S 3.5.1). 

The eastern side of the southern chimney is defined by the extension of the Radio Arc down to $b\sim-1^\circ$. Also here, the morphology of the Radio Arc changes from an array of thin filaments to a fuzz of emission, consistent with a dominant magnetic field with degrading flux density away from the plane, as a consequence of the diverging magnetic field with latitude (Fig. \ref{MeerKAT} and \ref{xmmMK}). 

The entire southern chimney appears to be filled with hot plasma and surrounded by radio emission. This is even more evident at its southernmost extension (Figs. \ref{xmmMK}, \ref{Molinari} and \ref{all}). Again this is easily understood as the aftermath of an outflow carving its way out and accumulating ISM material at the edges. We note that hot X-ray emitting plasma leaks beyond the southernmost radio edge. As for the northern cap, this implies that material at the edge is likely only partially impeding the observed outflow. 
Additional support for the "escaping hot plasma" scenario comes from the detection of recombination lines in its spectrum at this location (Nakashima et al. 2013). Those authors estimate an expansion time-scale on the order of $\sim8\times10^4$~yr, with an upper limit of $\sim1.1\times10^5$~yr (Nakashima et al. 2013). We note that recombining plasma is often associated with plasma expanding on a time-scale shorter than that required to reach ionization equilibrium. Therefore, the claimed presence of recombining plasma could be readily understood if some portion of the outflow were unimpeded in its expansion by the radio shell.  

\subsection{High-latitude hot plasma: Further evidence for a multi-epoch outflow}
\label{highb}

We observe that both the northern and southern chimneys merge at high latitudes with cooler, X-ray-emitting plasma, which is consistent with the X-ray emission observed in the \rosat\ map and attributed to the plasma around the edges of the \fermi\ bubbles (Ponti et al. 2019). 

The physical conditions of the high-latitude hot plasma and of the chimneys support the idea that the chimneys represent the channel replenishing the \fermi\ bubbles with energy and particles and that the current detailed morphology of the chimneys reflects the most recent episodes of energy injections. 

\subsection{Apparent association of the most prominent nonthermal radio filaments with the GC outflow}
\label{fila}

Since their discovery in the eighties, it has been suggested that the nonthermal radio filaments are tracers of an intense (dominating), pervasive and vertical magnetic field with an intensity of $\sim1$~mG (Morris \& Serabyn 1996). The nonthermal filaments would then appear anywhere there is a source of relativistic particles that illuminate the tube of field lines in which they are trapped (Morris \& Serabyn 1996). The recent discovery of groups of nonthermal filaments spatially organized to resemble "harps," provides considerable credibility to such an interpretation (Thomas et al. 2020). Several other interesting processes have also been invoked to explain the origin of the filaments (Lesch \& Reich 1992; Serabyn \& Morris 1994; Rosner \& Bodo 1996; Shore \& LaRosa 1999; Bicknell \& Li 2001; Yusef-Zadeh 2003; 2019; Bykov et al. 2017; Sofue 2020a).

{\it Does the GC outflow play a role in creating the nonthermal filaments?}
We note that the overlay of the \wise\ and \meerkat\ maps with the location of the strong shock suggested by Uchida et al. (1994) shows that the bright nonthermal filament C16 might originate at this location, although it could be coincidental (Figs. \ref{MeerKAT}, \ref{xmmMK} and \ref{all}). Indeed, in theory, strong shocks are expected to enhance the magnetic field strength and to accelerate particles, the two ingredients necessary to illuminate the filaments. 

In agreement with this scenario, we note that a large fraction of the nonthermal filaments are indeed observed either within or just outside of the chimneys (Fig. \ref{FC}). This raises the possibility that a fraction of these magnetic filaments might have been produced in a strong-field ($B\sim0.1-1$ mG) boundary zone surrounding the chimneys, perhaps by the compression of the GC magnetic field at that interface by the high gas pressure associated with the outflow. Alternatively, some filaments might be due to the (subsonic) stretching of a buoyant fluid element that will stretch the magnetic field, creating adjacent field lines of opposite polarity in its wake, which then gives rise to magnetic reconnection that accelerates particles to sufficient energies to shine via synchrotron emission in the radio band (see e.g., Churazov et al. 2013). Of course, if the nonthermal filaments are distributed within a cylindrical region surrounding the chimneys, then we should expect that some of them would be projected toward the interior of the chimneys as is observed. 

\section{Emerging picture}

Despite the complexity of the region --where projection effects are certainly relevant, where several competing physical effects might be at play, and where obscuration limits the breadth of our observing windows-- a picture appears to emerge from this multi-wavelength study. 
Indeed, the comparison of the X-ray, radio and infrared emission of the chimneys demonstrates the high level of interconnection of the emission in these various bands. Indeed, the same coherent features, extending for hundreds of parsecs, can be followed in all three bands. This demonstrates that the main players traced in each band, which are the hot plasma in X-rays, the warm dust in infrared and the shocks and nonthermal filaments in the radio, are interacting and deeply affecting each other. This strengthens the idea that they are the byproduct of the same phenomenon that drives them all. 

These multiwavelength data support the idea that the chimneys represent the channel connecting the quasi-continuous, but intermittent activity at the GC with the base of the \fermi\ bubbles. 
The prominent edges observed in the radio band are then the signposts of the most powerful and recent outflow from the central parsecs, which has created the radio shell partially filled with X-ray emitting plasma and swaths of warm dust at the boundary. 

\textbf{\textit{On the relation between the activity currently observed at the GC and the outflow of hot plasma which shaped the chimneys.}} \\
As shown by Ponti et al. (2019), the steep pressure gradient within Sgr~A's bipolar lobes indicates that there is an outflow of hot plasma within the central tens of parsecs. However, this small-scale hot outflow is not the driver of the outflow reaching the top of the chimneys. 

On the northwestern side of the chimneys, we observe that the pressure associated with the molecular outflow created a strong shock possessing a pressure that is about two orders of magnitude larger than the thermal pressure of the hot plasma (if volume filling) within the chimneys\footnote{Unless the hot outflow is moving at supersonic speeds ($v\gg500$~km~s$^{-1}$), also the ram pressure of the hot outflow is one-to-two orders of magnitude smaller than the pressure of the molecular outflow. }.
Again, on the northeastern side, the pressure appears to be dominated by the vertical magnetic field with a field strength of $\sim0.1-1$ mG. These comparisons suggest that either the hot outflow is supersonic or the chimneys were inflated and sculpted by an entity that is different from the currently observed hot plasma. 

In the latter case, the observed hot plasma might trace only a small fraction of the energy of the more powerful outflow which would be the driver of the observed phenomenology. 
Several candidates, including cosmic rays (Breitschwerdt et al. 1991; Yusef-Zadeh et al. 2019), Alfven or MHD waves (Sofue 2020a,b), fast-and-cold outflow resulting after rapid adiabatic expansion (Chevalier \& Clegg 1985; Heckman et al. 1990; Suchkov et al. 1994; Krumholz et al. 2017), very hot plasma, among others have been proposed. 
We stress that the presence of a fast-and-cold outflow is often invoked in starburst galaxies, where the initially hot plasma is rapidly adiabatically expanding, therefore transforming into a fast and cold flow (Heckman et al. 1990; Suchkov et al. 1994; Krumholz et al. 2017).
On the other hand, if indeed an enhancement of very hot plasma ($kT\sim7-10$~keV) is present within a degree of the Galactic plane at the GC, then this might be the driver of the outflow (Yamauchi et al. 1993; 2008; Koyama et al. 1986; but see Revnitsev et al. 2009). 

\textbf{\textit{The "relic" outflow scenario.}} \\
Alternatively, the outflow might have been significantly more powerful in the past, while its pressure has now dropped by orders of magnitude, although some pressure is still maintained by continued, lower-level activity. In such a scenario, hot plasma (with a significantly higher pressure than the one observed today) was previously flowing through the chimneys, then leaving the chimneys (on a sound crossing time of $t\sim3\times10^5$~yr) to energize the \fermi\ bubbles. The currently observed hot plasma within the chimneys would still be radiating because of its long cooling time and it would likely be in hydrostatic equilibrium with the Galactic potential. In this scenario, the relic hot plasma within the chimneys would currently drive no strong shocks. Therefore, in agreement with the multiwavelength observations, we would observe strong shocks at the boundaries of the chimneys only where they are driven by the slower moving phases of the GC outflow (i.e., \af), which are still tracing the older, more powerful activity. Additionally, in this relic outflow scenario, the high magnetic field strength (with $\sim0.1-1$ mG) might have been close to equipartition with the much more powerful past outflow. 
This picture appears in line with an active galactic nucleus-like type of activity, where the power of the outflow can change by orders of magnitudes on a relatively short time-scale. 
Clearly, a lower level of ongoing nuclear activity would still be required to produce features such as Sgr A's bipolar lobes, etc. 
In all of these scenarios, the formation of the northern cap and southern partially "enclosing" radio emission is likely associated with the bursty nature of the source powering the chimneys. 

\textbf{\textit{Interchanging dominant terms in the pressure balance and the role of the magnetic field.}} \\
The vertical magnetic field, present in the diffuse component of the GC, is likely to play an important role in collimating the outflow from the plane. 
At the same time, the compression and stretching of the fluid, induced by the outflow at the edges of the chimneys, will enhance the magnetic field, likely facilitating the generation of nonthermal filaments. 
The toroidal distribution of molecular material at the GC (Fig. \ref{Molinari}) must also have played a role in collimating the outflow. Indeed, it has been suggested that the $\sim5\times10^7$~M$_\odot$ of molecular mass within the central $\sim200$~pc of the plane (Dahmen et al. 1998; Molinari et al. 2011) might be able to redirect out of the plane even a powerful active galactic nucleous outflow (Zubovas et al. 2011). 

The total pressure driving the outflow is likely composed of several contributors. The data suggest that the magnetic pressure might be the dominant term in several places at the GC. Indeed, we ascribe the morphological difference between the eastern and western sides of the chimneys to a difference in the balance between the dominance in the pressure of matter and of magnetic field. 

\textbf{\textit{Testing these scenarios.}} \\
Future radio and X-ray observations of the GC will soon allow an assessment of the global framework proposed here and will provide insights to some of the many questions left open by this work. The most immediate question relates to the true location of the northwestern edge of the chimney. If this edge were located just in front of the GC as we have assumed, it would be consistent with the scenario proposed here. Were it instead located in the foreground Galactic disk, then it would have serious implications for the coherence of the north-south chimneys, and radio and infrared emission from the vicinity of the \af\ feature would then need to be sorted out from that arising in the foreground.

Additionally, high-resolution radio polarization and spectral index observations should eventually yield a much clearer distinction between thermal and nonthermal emission regions, thereby pointing to where the magnetic field plays an important role. Also, a more detailed  X-ray and radio investigation of the nature of the Protrusion feature is needed to determine whether it is an offshoot of the northern chimney, perhaps created by some directed energy flow from \sgras, or an unrelated but superimposed plasma volume created by supernovae in the central molecular zone. Finally, many fundamental things will be learned about the GC outflow when future X-ray spectrometers (such as \athena) can provide information on the velocity field of the hot plasma. 

\section{Acknowledgments}

This investigation is based on observations obtained with XMM-Newton, an ESA science mission with instruments and contributions directly funded by ESA Member States and NASA. This research has made use of the NASA/IPAC Infrared Science Archive, which is funded by the National Aeronautics and Space Administration and operated by the California Institute of Technology. 
This project acknowledges funding from the European Research Council (ERC) under the European Union’s Horizon 2020 research and innovation programme (grant agreement No 865637).

\begin{appendix}
\section{Comments on the recent suggestion that the northwestern edge of the chimney is a foreground feature}
\label{appendix}

As discussed in \S 3.2, the line of sight distance toward the northwestern edge of the chimney is still highly debated. 
Indeed, recently several authors have proposed that the northwestern edge of the chimneys is a foreground feature located within $\sim5-6$~kpc from us (Nagoshi et al. 2019; Tsuboi et al. 2020). 
These claims have been formulated on the basis of four points: \\
{\bf First: } The surprisingly low radial velocities (from $-4$ to $+10$~km~s$^{-1}$; Law et al. 2009; Nagoshi et al. 2019) of the radio recombination lines (RRL) along the edges of the northern chimney; \\
{\bf Second: } The observation of the silhouette of the northern chimney against low-frequency emission; \\
{\bf Third: } The observation of H$\alpha$ emission associated with the top of the northwestern chimney; \\
{\bf Fourth: } The distribution of the visual extinction from the Digitilized Sky Survey. \\
Indeed, Tsuboi et al. (2020) proposed that the northwestern edge of the chimney corresponds to a giant H {\sc ii} region in the Galactic disk. 
Additionally, Wang (2020) pointed out that the ionized plasma inside such an H {\sc ii} region might be the source of the scattering layer observed toward many GC sources, most notably the GC magnetar (Mori et al. 2013; Rea et al. 2013; Eatough et al. 2013; Bower et al. 2014). 

While the case for a location of the northwestern edge of the chimney in front of the main emission from the GC appears solid, its location at only few kiloparsecs from us leaves a few open issues.\\
{\bf First: } The first and most striking is related with the good degree of north-south symmetry of the chimney. In particular, the western edge of the chimney can be traced very well both in the X-ray and radio maps all the way from $b\sim-1.5$ to $\sim0.8^\circ$ (Fig. \ref{xmmMK}). 
If the northwestern edge of the chimney is located only a few kiloparsecs away, the striking continuity of the northern and southern chimneys would likely imply that the southern counterpart is also relatively local, and the challenge becomes explaining such a morphologically peculiar feature.
In that case, it is not obvious what could produce such an energetic feature with such a unique morphology, and it would have to be a truly remarkable coincidence that such a feature is centered on the GC. \\
{\bf Second: } The northwestern edge of the chimney appears intimately related with \af. \af\ is the most prominent part of a molecular cloud moving at very high positive velocities that strongly indicate its GC location.  
\af\ cannot be placed anywhere but in the central molecular zone (Serabyn \& Morris 1994). Its large positive velocity ($v\gg100$~km s$^{-1}$) has no known counterpart in foreground material, and there is no obvious mechanism to accelerate an individual cloud from the velocity range previously observed in this direction (from $-60$ km s$^{-1}$ to $+20$ km s$^{-1}$) up to such a high positive velocity.  \\
{\bf Third: } The northeastern side of the chimney is nestled up against the radio Arc and its extended filaments and polarized plumes, which are clearly located at the GC. \\

We note that one of the most relevant arguments to place the northwestern side of the chimney at few kpc from us are the low radial velocities of the RRL. 
We note that RRL can be inverted and produce weak maser amplification and that the low density and temperatures in the surface layers of foreground molecular clouds in the Galactic disk could produce such inversions. In this scenario, the radial velocities of the RRL are characteristic of the foreground clouds, therefore they do not necessarily imply that the northwestern edge of the chimney must be located in the Galactic disk. We leave the detailed discussion of this scenario to a future publication (Morris et al. in prep.). 

The observation of the silhouette of the northern chimney against low-frequency emission definitely places most of the northern chimney in front of the bulk of the low-frequency radio continuum produced at the GC. We propose that both the northern and southern chimneys are rooted at the GC, but they are significantly tilted toward us, such that the northern chimney is in front of the bulk of the radio emission arising in the GC.  

The Galactic latitudes at which H $\alpha$ is seen are relatively large, so if the GCL is at the GC, the extinction at those latitudes might not be so large that the H $\alpha$ line would be completely extinguished. Therefore, the presence of H $\alpha$ is not a convincing argument for the foreground hypothesis (see Fig. 5 of Law 2010).

While these arguments do not address all the open issues (e.g., the visual extinction toward the northwestern edge, which supports the hypothesis that the northwestern edge of the chimneys is located in the foreground), we assume that the chimney is a coherent structure, with a significant tilt, and rooted at the GC. However, we caution the reader that some portion of the features that in projection appear superimposed on the chimney might have a different origin. 

\end{appendix}

\end{document}